\documentclass{jfm}

\usepackage{graphicx}
\usepackage{amsmath}
\usepackage{epstopdf}
\usepackage{amssymb}
\renewcommand{\Re}{\Rey}
\newcommand{\Ta}{T\!a}
\newcommand{\We}{W\!e}

\newcommand{\Nuw}{N\!u_{\omega}}

\newcommand{\DR}{D\!R}
\usepackage[colorlinks = true,allcolors = {blue}]{hyperref}

\shorttitle{The influence of wall roughness on bubble drag reduction in TC turbulence}
\shortauthor{R. A. Verschoof, D. Bakhuis, P.A. Bullee, S. G. Huisman, C. Sun, D. Lohse}

\title{The influence of wall roughness on bubble drag reduction in Taylor-Couette turbulence
}
\author{Ruben A. Verschoof\aff{1}, 
Dennis Bakhuis\aff{1}, 
Pim A. Bullee\aff{1}, 
Sander G. Huisman\aff{1},
Chao Sun\aff{2,1}
 \corresp{\email{chaosun@tsinghua.edu.cn}},
 \and {Detlef Lohse}\aff{1,3}
 \corresp{\email{d.lohse@utwente.nl}}
}
\affiliation{
\aff{1}Physics of Fluids, Max Planck Institute for Complex Fluid Dynamics, MESA+ Institute and J. M. Burgers Center for Fluid Dynamics, University of Twente, P.O. Box 217, 7500 AE Enschede, The Netherlands
\aff{2}Center for Combustion Energy and Department of Thermal Engineering, Tsinghua University, 100084 Beijing, China
\aff{3}Max Planck Institute for Dynamics and Self-Organization, Am Fassberg 17, 37077 G\"{o}ttingen, Germany}

\begin{document}
\maketitle

\begin{abstract} 
We experimentally study the influence of wall roughness on bubble drag reduction in turbulent Taylor-Couette flow, i.e.\ the flow between two concentric, independently rotating cylinders. We measure the drag in the system for the cases with and without air, and add roughness by installing transverse ribs on either one or both of the cylinders. For the smooth wall case (no ribs) and the case of ribs on the inner cylinder only, we observe strong drag reduction up to $DR=33\%$ and $DR=23\%$, respectively, for a void fraction of $\alpha=6\%$. However, with ribs mounted on both cylinders or on the outer cylinder only, the drag reduction is weak, less than $DR=11\%$, and thus quite close to the trivial effect of reduced effective density.
Flow visualizations show that stable turbulent Taylor vortices --- large scale vortical structures --- are induced in these two cases, i.e. the cases with ribs on the outer cylinder. These strong secondary flows move the bubbles away from the boundary layer, making the bubbles less effective  than what had previously been  observed for the smooth-wall case.  Measurements with counter-rotating smooth cylinders, a regime in which pronounced Taylor rolls are also induced, confirm that it is really the Taylor vortices that weaken the bubble drag reduction mechanism. Our findings show that, although bubble drag reduction can indeed be effective for smooth walls, its effect can be spoiled by e.g.\ biofouling and omnipresent wall roughness, as the roughness can induce strong secondary flows. 
\end{abstract}


\section{Introduction}
In the maritime industry, air lubrication is seen as one of the most promising techniques to reduce the overall fuel consumption \citep{kod00,Foeth2008,Makiharju2012}. Air lubrication has been studied for several decades, and it is found that a few percent of air can significantly decrease the overall friction, e.g. with 4\% bubbles, drag reductions up to 40\% were shown \citep{gil13, ver16}.  Notwithstanding its clear industrial potential, it remains difficult to translate highly controlled laboratory results to drag reduction  in large-scale vessels. So far, the vast majority of studies on bubble drag reduction (DR) have been performed in test facilities with purified water and smooth walls, see e.g.\ the review articles by \citet{cec10} and \citet{murai2014}.
However, many surfaces in industry are rough to some extent, and also initially smooth surfaces can become rough by means of corrosion, cavitation, mineral scaling, and (bio)fouling \citep{sch07,Schultz2011}. Furthermore, the dynamics of bubbles are strongly affected by any dissolved ions in oceanic water and surfactants \citep{takagi2008,takagi2011}. 
As the conditions in controlled experiments and real applications are so much different, one can expect that bubble DR experiments will lead to very different results in practice. 
Only a limited number of studies focussed on `non-ideal' DR, either through  wall roughness \citep{deu04,ber07,elbing2008}, surfactants or seawater \citep{Takahashi2001,Winkel2004,Shen2006,elbing2008,ver16}, and their results are somewhat inconsistent. Some studies found that wall roughness completely eliminates any drag reduction \citep{ber07}, whereas others show that roughness does not affect, or even enhance, drag reduction \citep{deu04,elbing2008}. Therefore, there is a clear need to better understand the influence of wall roughness on bubble drag reduction.

In this work, we study the effect of wall roughness on bubble drag reduction. To do so, we employ the Taylor-Couette (TC) system, i.e.\ the flow between two concentric, independently rotating cylinders \citep{far14,gro16}. TC flow is one of the canonical flow systems in which fluid mechanics concepts and theories are tested. Among the advantages of using a TC setup are the ease with which the global void fraction $\alpha$ is controlled, the absence of any streamwise spatial transients, and as it is a closed system, an exact balance that  connects the global torque measurements with the local energy dissipation rate. The driving and response of the system are characterized by the Taylor number $\Ta$ and the Nusselt number $\Nuw$, respectively \citep{gro16}. The Nusselt number is defined as the ratio of the convective momentum transport to the diffusive flux, and using it underlines the close analogy between Taylor-Couette flow and Rayleigh-B\'enard convection \citep*{eck07b}. In the currently studied parameter regime, an effective scaling of $\Nuw \propto \Ta^{0.4}$ is observed \citep{gil11}.
 The TC setup has been used frequently to study (bubble) drag reduction in turbulent flows \citep{ber05,gil13,ver16,ros16,sar16}, even numerically \citep{sug08b,spa17}. In these studies, it was shown that a small air fraction can considerably reduce the drag. E.g.\ with a void fraction of $\alpha=4\%$, a drag reduction of $40\%$ was observed \citep{gil13,ver16}, which is significantly larger than the trivial effects of affected effective density and viscosity. These studies highlighted the importance of bubble deformability for large drag reduction, and thus a sufficiently large Weber number $\We = \rho D_b u^2/\sigma $, in which $\rho$ is the fluid density, $D_b$ the bubble diameter, $u$ a characteristic velocity and $\sigma$ the interfacial surface tension. It was shown that large Weber number bubbles, i.e.\ large and deformable bubbles, are crucial to efficiently reduce the drag.

Wall roughness, on the other hand, obviously increases the friction. Its effects have been studied extensively for single-phase turbulence flows, mostly in pipe  or channel flow configurations given their industrial relevance, see e.g.\ \citet{mar10b,fla14} and references therein.  For TC flow, adding sufficiently large, rough ribs results in a $\Nuw \propto \Ta^{1/2}$ scaling, rather than the aforementioned $\Nuw \propto \Ta^{0.4}$ smooth wall scaling \citep{cad97,ber03,zhu18}. The $\Nuw \propto \Ta^{1/2}$ scaling, mathematically equivalent to a constant friction coefficient in the fully rough regime, is the mathematical upper bound to the transport of momentum. 
In this regime, the roughness decreases the near wall velocity gradient, whereas the streamwise velocity fluctuations are increased. The bubble dynamics are largely governed by the motion of the surrounding fluid, but to which extent any drag reduction is affected by the changed fluid motion is  unknown.

A number of studies focussed on wall modifications to stimulate air to attach to the inner cylinder wall of Taylor-Couette flow, either by a hydrophobic coating \citep{sri15} or by using cavitors to try to create an air layer \citep{Verschoof2018}.    \citet{ber07} studied the effects of roughness on bubble drag reduction, and found that ribs attached to both cylinders prevent bubbles from reducing the overall friction. Therefore, it was suggested that bubbly drag reduction is a boundary layer effect. However, the exact reason why the drag reduction was lost remained elusive. Therefore,  here we aim to repeat and extend those experiments in a more accurate and controlled setup and to visualize the flow, to better understand the physics of the aforementioned conclusions.

\section{Experimental method}
%\section{Torque}
\begin{figure}
\centering
\includegraphics[width = 1.0\linewidth]{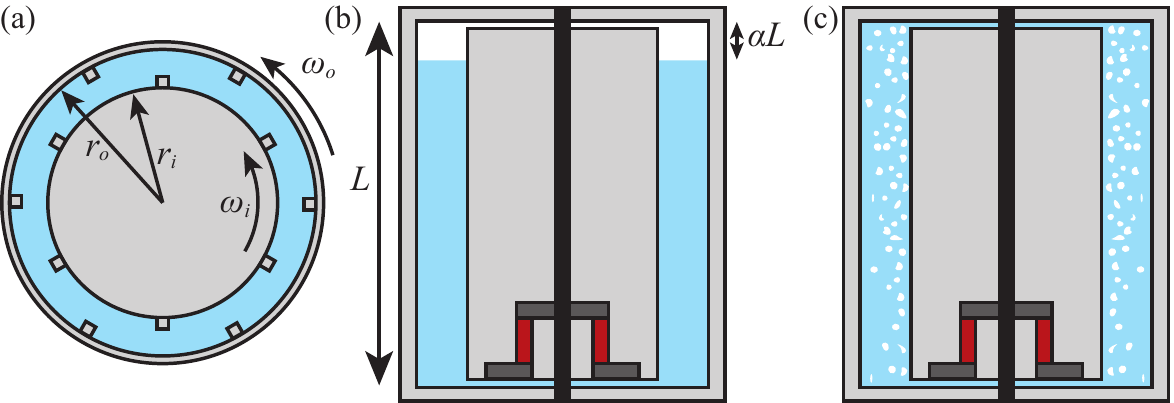}
\caption{{\bf Experimental setup}.  (a) Top view schematic of the T$^3$C facility. We attached 6 vertical transverse ribs (not to scale) equally distributed around the perimeter of the inner cylinder, the outer cylinder, or both cylinders. We also measure a smooth-wall case without any ribs. (b) Vertical cross-section of the setup at rest, showing the position of the torque sensor. To control the void fraction, we fill the cylinder only partially with water, so that the void fraction $\alpha$ is controlled by measuring the relative height of the water level. (c) Vertical cross-section of the setup during a measurement. The free surface disappears, and all air is entrained by the turbulent flow (bubbles not to scale). }
\label{Chap_Seven_fig:setup}
\end{figure}
The experiments are performed in the Twente Turbulent Taylor-Couette facility (T$^3$C) \citep{gil11a}, in which the flow is fully turbulent. TC flow is driven by the angular velocity of the inner and outer cylinder, denoted by $\omega_i$ and $\omega_o$, respectively.
The setup has a height of $L=927$ mm, an inner radius $r_i=200$ mm, an outer radius $r_o=279.4$ mm, giving a gap width $d=r_o-r_i=79.4$ mm. The geometry can therefore be described by two geometric parameters; the radius ratio $\eta=r_i/r_o=0.716$  and the aspect ratio $\Gamma = L/d = 11.7$, see also figure \ref{Chap_Seven_fig:setup}a. 
The inner cylinder and outer cylinder rotate up to $f_i=\omega_i/(2\pi) =  10$ Hz and $f_o = -4$ Hz, respectively. 
These result in two Reynolds numbers: $\Re_{i,o} = \omega_{i,o} r_{i,o} (r_o-r_i)/\nu$, respectively, in which $\nu$ is the viscosity of the working fluid, and a rotation ratio $a=-\omega_o/\omega_i$. We here express the driving using the Taylor number $\Ta = [(1+\eta)^4/(64\eta^2)]d^2 (r_i+r_o)^2( \omega_i-\omega_o)^2\nu^{-2} \propto (\Re_i - \eta \Re_o)^2$, which thus incorporates the rotation of both cylinders in one dimensionless number. In the current study, we measure at Taylor numbers of $\mathcal{O}(10^{12})$, or, equivalently, Reynolds numbers up to $\mathcal{O}(10^6)$. 
The primary response parameter is the torque $\tau$ necessary to maintain the inner cylinder at a constant angular velocity. The torque is measured with a co-axial torque transducer (Honeywell Hollow Reaction Torque Sensor 2404-1K, maximum capacity of 115 Nm), which is located inside the inner cylinder to avoid measurement errors due to seal and bearing friction, see figure \ref{Chap_Seven_fig:setup}b. The torque is made dimensionless with the torque for laminar non-vertical flow, resulting in the Nusselt number: $\Nuw = \tau / \tau_{lam}$, with $\tau_{lam} = 4 \pi L \rho \nu r_i^2 r_o^2 ( \omega_i-\omega_o) / (r_o^2 -r_i^2) $. 
The flow is cooled through both endplates to counteract viscous heating, keeping the water temperature constant within $T=21 \pm0.5^{\circ}$C. Although the effective viscosity and density are altered by the presence of bubbles, we chose to consequently use the pure water material properties, as we are interested in the net changes in drag.

The cylinders are made rough by attaching 6 transverse ribs over the entire height of the cylinders, as shown in figure \ref{Chap_Seven_fig:setup}a. The rib dimensions are $6$ mm by $6$ mm, corresponding to 7.5\% of the gap width, and  to $\mathcal{O}(10^3)$ in wall units, depending on the Taylor number and roughness case. We study the torque and resulting drag reduction for 4 cases: both cylinders smooth (SS), both cylinders rough (RR), and roughness on either only the inner cylinder (RS) or only on the outer cylinder (SR). We here chose to apply rib roughness, which, for the RR case, causes the flow to be in the ``fully rough'' state, or the ``asymptotic ultimate turbulence'' regime \citep{zhu18} in the studied parameter regime. In this regime, the behaviour in the boundary layers becomes independent of the viscosity. Consequently, a $\Nuw \propto \Ta^{1/2}$ is observed rather than the effective $\Nuw\propto \Ta ^{0.4}$ scaling found for the smooth wall case in the currently studied parameter regime \citep{kra62,gil11,zhu18}. For the cases of ribs on a single cylinder, the exponent $\gamma$ of the $\Nuw \propto \Ta^{\gamma}$-scaling is between these two bounds.

The gap is either partially or completely filled with water, so that the void fraction is precisely set between $0\%\leq \alpha \leq 6\%$, see figure \ref{Chap_Seven_fig:setup}b. We determine the void fraction with both cylinders at rest. During a flow measurement, the air is distributed over the height of the cylinder because of turbulent mixing, see figure \ref{Chap_Seven_fig:setup}c. We note that a perfect homogeneous axial distribution is not feasable, even with continuous bubble injection through the bottom end cap \citep{gil13}, but it becomes more homogeneous with increasing Taylor number.

\section{Results}
\begin{figure}
\centering 
\includegraphics[width = 0.8\linewidth]{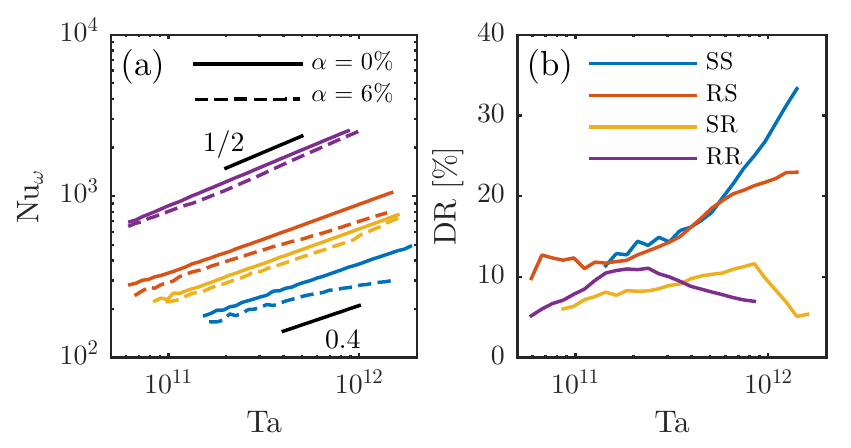}
\caption{(a) Dimensionless angular velocity flux $\Nuw$ as a function of $\Ta$ for $\alpha=0\%$ and $\alpha=6\%$. To increase the readability, we do not show $\Nuw$ for $\alpha=\{2\%,4\%\}$, which are used to calculate the DR shown in figure \ref{Chap_Seven_fig:torque}. The two short black lines indicate  the $\Nuw\propto\Ta^{\gamma}$ scaling relations for the pure liquid cases. The exponents are $\gamma=0.4$ and $\gamma=1/2$ for the SS and RR cases, respectively. {(b)} Resulting drag reduction as a function of $Ta$. The outer cylinder is stationary. The DR is calculated with equation (\ref{eq:DR}).}
\label{Chap_Seven_fig:torque_all}
\end{figure}
\begin{figure}
\centering 
\includegraphics[width = 0.8\linewidth]{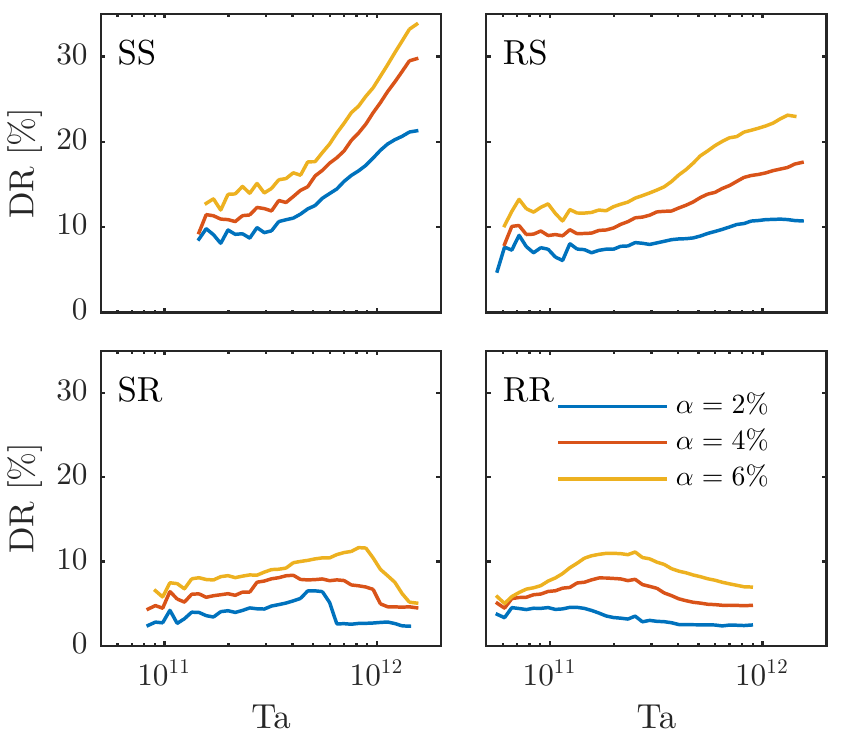}
\caption{Drag reduction percentages as a function of $\Ta$ for all roughness cases. In the RR case, the highest achievable Taylor number is slightly smaller due to experimental limitations. The outer cylinder is stationary. The DR is calculated with equation \ref{eq:DR}.}
\label{Chap_Seven_fig:torque}
\end{figure}

We measure the torque and  present our findings in figure \ref{Chap_Seven_fig:torque_all} and figure \ref{Chap_Seven_fig:torque}. The drag reduction is calculated as 
\begin{equation}
\DR = 1- \frac{\Nuw(\alpha)}{\Nuw(\alpha=0)},
 \label{eq:DR}
\end{equation}
 in which we compare the $\Nuw$ values for the same roughness case. In figure \ref{Chap_Seven_fig:torque_all}, we show the Nusselt number and resulting drag reduction for all roughness cases. As was shown before, $\Nuw$ depends tremendously on the applied roughness \citep{zhu18}.  In the current study, however, we are more interested in the {\it relative} bubbly drag reduction as compared to the smooth wall case, rather than the absolute friction increase by roughness. From figure  \ref{Chap_Seven_fig:torque_all}  two different regimes can be distinguished: we observe strong drag reduction for the SS and RS cases --- up to DR $=33\% $ with a void fraction of $\alpha=6\%$, whereas for the SR and RR cases, the DR is only weak --- with the same void fraction never exceeding DR $= 12\% $ .

To further study the DR per roughness case, we show the drag reduction for void fractions of $2\%$, $4\%$ and $6\%$ in figure \ref{Chap_Seven_fig:torque}. The DR increased monotonically with increasing void fraction for all cases. In the weak drag reduction cases (SR and RR), the DR is quite close to the trivial effect of reduced global density, which equals $\rho_{\text{eff}} = \rho(1-\alpha) + \alpha \rho_{\mathrm{air}} \approx \rho(1-\alpha)$, in which $\rho_{\mathrm{air}}$ is the air density.  For the RS and SS cases however, the drag reduction is significantly larger than the reduced density effect. Interestingly, given the strong DR in the RS case, it is clear that wall roughness does not necessarily prevents strong bubble drag reduction. For both the RS and SS cases, the drag reduction increases with Taylor number \citep{gil13}, contrasting the SR and RR cases, in which the drag reduction does not have a clear monotonic Taylor number dependence. 

%\section{Visualizations}
\begin{figure}
\centering
\includegraphics[width=\textwidth]{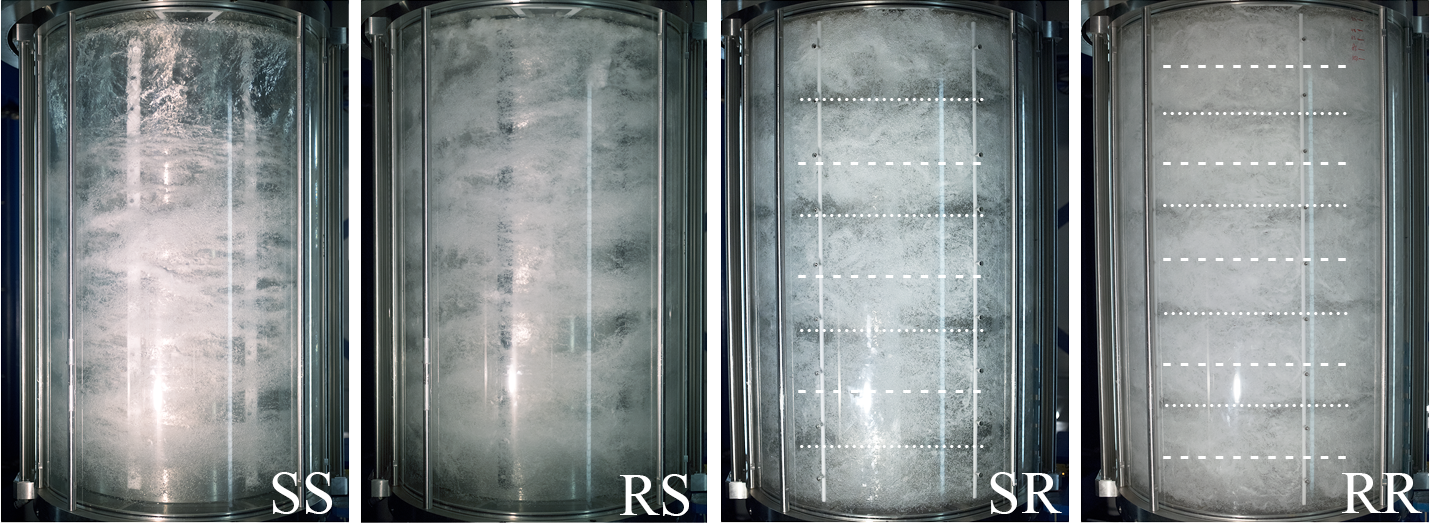}
\caption{Instantaneous photographs of the flow for all four roughness cases: SS, RS, SR and RR. Clear differences in the flow patterns are visible. In the SS and RS cases, we see turbulent streaks, but no stable structures. Clear stable Taylor rolls are visible for the SR and RR cases. We indicate the position of the rolls by the dashed line, and indicated the roll pairs by the dotted line. The Taylor number is $\Ta= 1.5\times 10^{12}$, except for the RR case ($\Ta = 8.4\times 10^{11}$), while the outer cylinder is kept stationary. The void fraction in all cases is $\alpha=6\%$. Note that in all cases the bubbles are not homogeneously distributed  over the height, this is most visible in the SS case.}
\label{Chap_Seven_fig:visu}
\end{figure}

To better understand the flow dynamics, we visualize the flow for a void fraction of $\alpha=6\%$. As shown in figure \ref{Chap_Seven_fig:visu}, for all four cases the flow structures are significantly different. In the SS and RS cases, clear streaks and patterns are visualized by the bubbles, but stable turbulent Taylor vortices are not observed. For both cases with ribs on the outer cylinder, i.e.\  the SR and RR cases, we do however observe stable turbulent Taylor vortices. 

The existence and the dynamics of Taylor vortices have been studied extensively for the single-phase smooth-wall case  \citep{lat92a,lat92,lew99,hui14,vee16b,gro16}. In the explored Taylor number regime, measurements showed that for sufficiently strong turbulence ($\Re_i>10^5$),  stable Taylor rolls do not exist in the pure inner cylinder rotation regime, and are only present in the counter-rotating regime \citep{ost14pd,gro16}.  Roughness elements promote the ejection of turbulent plumes, leading to localised radial flows towards the outer cylinder  \citep{zhu16,top17}. As the TC system is closed, consequently a radial flow towards the inner cylinder must be present. These flows can organise themselves as stable Taylor rolls. Thus, as the roughness promotes the ejection of turbulent plumes, the existence of Taylor vortices is stimulated.

The roll dynamics observed with wall roughness are different than what has been observed hitherto in the same setup. Earlier studies found  6 or 8 rolls for the smooth-wall case with  counter-rotating cylinders \citep{hui14,vee16b}. Here, for pure inner cylinder rotation, we see 10 rolls for the RR case, whereas for the SR case we observe 8 rolls.
The number of rolls is related to the aspect ratio $\Gamma$, which depends on the gap width. The roughness elements decrease the `effective gap width', and thus increase the apparent aspect ratio $\Gamma_{\text{eff}}$. Therefore, the system allows for an increased number of rolls \citep{vee16b}

We argue that the existence of the Taylor vortices is the underlying mechanism through which the effectiveness of bubble drag reduction is reduced in the SR and RR cases. To effectively decrease the drag, it is crucial that large bubbles are present in or close to the boundary layer \citep{gil13,ver16,spa17}. The flow visualizations show that the bubbles are dragged away from the inner cylinder wall by low vorticity regions, here in the form of turbulent Taylor vortices. Therefore, as the bubbles do not accumulate close to the inner cylinder, the drag reduction almost vanishes, and becomes close to the trivial effect of the reduced global effective density, as was shown in our torque measurements.

%%%%%%%%%

\begin{figure}
\centering

  \includegraphics[width=1\linewidth]{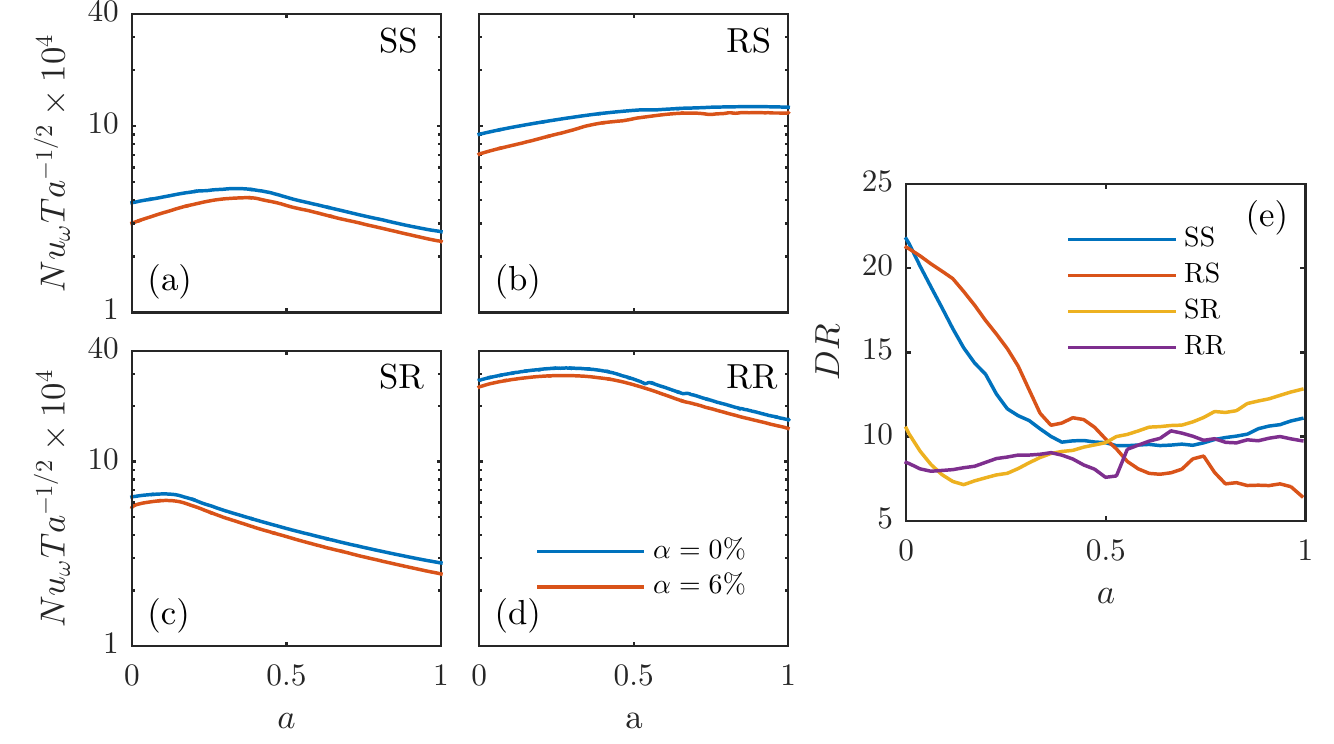}
\caption{(a-d) Dimensionless torque as a function of rotation ratio $a$ for the roughness cases SS, SR, RS, and RR with and without air.  The Taylor number is kept constant at $\Ta= 7.3 \times 10^{11}$. We compensate $\Nuw$ by $\Ta^{0.5}$ to remove viscosity changes due to temperature fluctuations, similar as in \citet{hui14}. (e) Resulting drag reduction as a function of rotation ratio $a$, calculated here as $\DR =  1 - \frac{\Nuw\Ta^{-0.4}(\alpha=6\%)}{\Nuw\Ta^{-0.4}(\alpha=0\%)}$.  }

\label{Chap_Seven_fig:torque_a}
\end{figure}

Up to here, we showed that in the SR and RR cases, the ribs induce turbulent Taylor vortices, and we argued that Taylor rolls eliminate DR. One could presume that the rolls, instead of being the underlying physical explanation, merely coincide with the weak DR. To further prove the effect of turbulent  Taylor vortices on bubble DR, we study the DR behavior in the counter-rotation regime. For the SS case, pronounced stable turbulent  Taylor vortices exist in the counter-rotating regime between approximately $0.1 \leq a \leq 0.5$ \citep{gil12,hui14, vee16b}.  By measuring the DR as a function of rotation ratio $a=-\omega_o/\omega_i$ while keeping the Taylor number constant, we can directly show the influence of Taylor rolls on the effectiveness of air lubrication. 
For all roughness cases, we show the torque in figure \ref{Chap_Seven_fig:torque_a}(a-d) and the resulting DR in figure \ref{Chap_Seven_fig:torque_a}(e). As already shown before, we observe strong DR for the SS and SR cases at $a=0$. Then, for increasing $a$, we see that the DR decreases. The observation is very similar to the above discussed weak DR in the SR and RR cases, namely in the counter-rotating regime the bubbles are trapped in the Taylor rolls, dragged away from the boundary layer and unable to effectively decrease the drag. For the SR case, the strength of the turbulent Taylor vortices decreases with increasing outer cylinder. Consequently, an increase in DR is observed. For the RR case, the DR remains weak for all cases, as the turbulent Taylor vortices exist for the entire scanned parameter space.

\section{Conclusions}
 
To conclude, we studied the influence of wall roughness on bubble drag reduction in a highly turbulent flow. We showed that in the SR and RR cases wall roughness promotes stable turbulent Taylor rolls, which induce strong secondary flows, suppressing the drag reduction. Bubbles are captured in low vorticity regions, and therefore dragged away from the inner cylinder boundary layer. As a result, the drag reduction is mostly lost, and the effective drag reduction is close to the trivial effect of reduced global density. 
These findings help us to understand earlier studies on air lubrication and wall roughness, which had conflicting results whether roughness influences bubble drag reduction. We here distinguish two different regimes: (i) a regime with strong drag reduction if the roughness does not introduce strong secondary flows. And, (ii) a regime with weak drag reduction if strong secondary flows are induced by the roughness.

Future work  includes  studies on  wall roughness combined with bubbles  in other types of setups, e.g.\ flat plates, or pipe flow. In these setups, roughness increases the velocity fluctuations but not necessarily induces stable large-scale  secondary flows, and thus the bubble DR behaviour might be significantly different than in the current study. Moreover, as we here limited ourselves to the influence of rib roughness, more work is needed to understand the influence of more realistic types of roughness.

\begin{acknowledgments}
We thank Tom van Terwisga (MARIN, TU Delft) for the continuous and stimulating collaboration over the years on drag reduction in the marine context. We also thank Dennis van Gils, Gert-Wim Bruggert, and Martin Bos for their technical support. The work was financially supported by NWO-TTW (project 13265). We acknowledge support from EuHIT and MCEC. Sun and Bakhuis acknowledge financial support from VIDI grant No. 13477, and the Natural Science Foundation of China under grant no. 11672156. Bullee acknowledges NWO-TTW (project 14504).

\end{acknowledgments}

\end{document}